# Strain selectivity of SiGe wet chemical etchants

M. Stoffel[1,*], A. Malachias[1], T. Merdzhanova[1]

F. Cavallo[1,2], G. Isella[3], D. Chrastina[3], H. von Känel[3]

A. Rastelli[1,2] and O. G. Schmidt[1,2]

[1] *Max-Planck-Institut für Festkörperforschung, Heisenbergstraße 1,*

*D-70569 Stuttgart-Germany*

[2] *Institute for Integrative Nanosciences, IFW Dresden, Helmholtzstrasse 20,*

*D-01069 Dresden, Germany*

[3] *L-NESS, Dipartimento di Fisica del Politecnico di Milano, Polo Regionale di Como,*

*Via Anzani 42, I-22100, Como, Italy*

Abstract

We investigate the effect of strain on the etching rate of two SiGe wet etchants, namely $NH_4OH:H_2O_2$ and $H_2O_2$. For both etchants, we found that there is no appreciable strain selectivity, i.e. the etching rates do not depend on the actual strain state in the SiGe films. Instead, for the $NH_4OH:H_2O_2$ solution, the rates are primarily determined by the Ge content. Finally, we show that both etchants are isotropic with no preferential etching of particular facets.

[*] corresponding author: **m.stoffel@fkf.mpg.de**

**1. Introduction**

Wet chemical etching techniques are widely used in semiconductor technology for device processing, for identifying crystal symmetries [1] or for revealing threading dislocations [1,2]. Etching techniques represent a critical step in the fabrication of novel micro-and nano-devices. For example, the definition of electrical contacts on a buried base layer requires the selective etching of the top emitter layer in a vertical transistor structure [3]. In addition, selective wet chemical etching was also successfully used for various applications, such as for fabricating microelectromechanical systems [4] or Si nanowires [5], for nanoscale patterning of Si/SiGe heterostructures in combination with electron beam lithography [6] and for engineering novel SiGe/Si or InGaAs/GaAs based micro-and nano-objects [7-9].

Recently, various etchants were also used in combination with atomic force microscopy (AFM) as a tool to investigate the composition of strained semiconductor islands (self-assembled quantum dots, QDs) or to reveal the morphology of QDs encapsulated in a semiconductor matrix. Based on the selectivity of a $H_2O_2$ solution that etches $Si_{1-x}Ge_x$ alloys with x>65% over pure Si [10], Denker et al. [11] provided evidence of a lateral composition profile in SiGe pyramids. Katsaros et al. [12] extended the previous approach to SiGe domes and suggested a kinetic model to account for the observed lateral profiles. Alonso et al. [13] further used the same techniques to investigate SiGe dislocated islands. Later on, Leite et al. [14] investigated the alloying mechanism in epitaxial SiGe nanocrystals using a $NH_4OH:H_2O_2$ solution. Wet chemical etching was also used to corroborate compositional results obtained independently from X-ray diffraction measurements [15,16] and to shed new light into the evolution of SiGe/Si(001) islands either during in-situ annealing [17] or during Si capping [18]. By using KOH to selectively remove the Si capping layer, the morphology of buried QDs was studied in detail [19,20] and a similar approach, based on a $NH_4OH:H_2O_2:H_2O$ solution, was successfully employed to investigate buried InGaAs/GaAs QD structures [21, 22].

Etching studies were also performed on polycrystalline SiGe films and the influence of different parameters such as etching temperature, doping concentrations, on the etching rates was discussed [23]. However, to the best of our knowledge, a systematic study of the effect of strain on the etching rate of solutions typically used for investigating the composition profiles of SiGe islands was not presented so far. However, since self-assembled QDs are strained, it is indispensable to determine whether and to what extent the etching rate depends on the strain state of the SiGe material. In order to address this open issue, we investigate planar SiGe structures having the same nominal composition but different strain states (relaxed, 20% tensile or 20% compressively strained). We consider here two SiGe etchants, namely $NH_4OH(1):H_2O_2(1)$ and $H_2O_2$. Both etchants are known to selectively etch $Si_{1-x}Ge_x$ alloys over pure Si. The former etches $Si_{1-x}Ge_x$ alloys with x>20 % [24] while the latter etches $Si_{1-x}Ge_x$ alloys with Ge contents higher than $(65 \pm 5)\%$ [10]. For the samples considered here, we find that both etchants are insensitive to strain, i.e. the etching rates do not appreciably depend of the strain state of the films but critically depend on the alloy composition. Finally, we show that both etchants are isotropic with no preferential etching of particular facets.

## 2. Experimental details

The samples used for this study were grown by low energy plasma enhanced chemical vapor deposition (LEPECVD) at substrate temperatures between 490°C and 740°C. The layer thicknesses and compositions were determined using secondary ion mass spectroscopy (SIMS) and showed excellent agreement with the nominal values. The strain state in the SiGe films was probed by X-ray coplanar diffraction in the vicinity of the Si(004) Bragg peak using a Bruker D8 Discover diffractometer equipped with an Eulerian cradle. Cu K$\alpha$ radiation emerging from the point focus (0.1 mm x 0.1 mm) of a rotating anode source operating at 50 kV and 20 mA was converted to a quasi parallel beam by a parabolically graded multilayer mirror (Xenocs). The size of the beam at the sample position was approximately 1 mm$^2$. The diffracted beam was detected by a scintillation counter. Prior to the SiGe etching experiments, we used photolithography to define mesa structures on the sample surface. In the next step, 20-100 nm of Cr were deposited followed by lift-off. The as-defined structures were then used as etch masks for the subsequent etching steps. The samples were dipped in either a 1:1 vol. (28% NH$_4$OH/ 31% H$_2$O$_2$) or in a 31% H$_2$O$_2$ solution (VLSI Selectipur, Merck). All etching experiments were performed at room temperature without stirring the solution. Before each etching step, the samples were dipped in a diluted hydrofluoric acid solution for 60 s to remove the native oxide layer. Immediately after etching, the samples were rinsed for 30 s in deionized water and subsequently dried with nitrogen gas. The etching rate was determined from the height profile measured using AFM after various etching times. In order to compare the etching rates, we measured the height profile at the *same surface location* and we prepared the etching solutions from the same constituents for all our experiments. All etching steps were performed within one month ensuring thus a good chemical stability of the etching solutions. In order to probe a possible etching anisotropy, manifesting in the formation of facets during etching, we defined circular, ring-like structures using photolithography and subsequent metal deposition. The deeply etched mesas were then characterized by scanning electron microscopy (SEM).

## 3. Results and discussion

The generic sample structure used for our experiments is shown in Figure 1*(a)*. First, a relaxed SiGe buffer consisting of either a compositionally-linearly graded layer, or a constant composition layer with final Ge fraction *y*, was grown on top of a Si(001) substrate. Then, a 50 nm thick $Si_{1-x}Ge_x$ film having a constant Ge fraction *x* is grown. *x* was chosen to be equal to *y* for relaxed films, while biaxially tensile/compressive strained films were obtained by choosing *x=y*-0.2 and *x=y*+0.2, respectively. Finally, all structures were capped with a 6 nm thick Si capping layer. In order to make the strained $Si_{1-x}Ge_x$ film accessible to the etching solution, the top Si layer was first removed in a 2M potassium hydroxide (KOH) solution. The latter solution is known to etch selectively Si over $Si_{0.80}Ge_{0.20}$ with a selectivity of about 100:1 [25], leaving thus the strained (or relaxed) SiGe films unetched. The Ge contents measured by SIMS for two particular samples which contain a 50 nm thick $Si_{0.40}Ge_{0.60}$ film grown either on top of a 1 μm thick $Si_{0.20}Ge_{0.80}$ relaxed buffer (tensile strained layer, sample A) or on a 7 μm thick $Si_{0.60}Ge_{0.40}$ graded layer (compressively strained layer, sample B) are shown in Figures 1*(b)* and 1*(c)*, respectively. In both cases, the Ge content is in excellent agreement with the nominal values and does not vary appreciably throughout the strained $Si_{0.40}Ge_{0.60}$ film. The layer thicknesses determined from the SIMS profiles agree also quite well with the nominal layer parameters within 10%.

In order to probe the strain state in the films, we measured X-ray longitudinal scans along the (004) specular direction for the samples A and B described above. Figures 2*(a)-(b)* show longitudinal (θ-2θ) scans spanning from the unstrained Ge(004) reciprocal space position ($q_r$= 4.44 Å$^{-1}$) to the bulk Si position ($q_r$= 4.63 Å$^{-1}$) for samples A and B after removal of the Si cap layer. The X-ray scans were performed before (solid dots) and after 7 min etching in a NH$_4$OH(1):H$_2$O$_2$(1) solution (open dots). From the measurements, one can draw two preliminary observations. First, the diffraction peak originating from the Si$_{0.40}$Ge$_{0.60}$ film is observed at $q_r$= 4.54 Å$^{-1}$ for sample A and at $q_r$= 4.49 Å$^{-1}$ for sample B. This is consistent with the out-of-plane lattice contraction (expansion) of a biaxially strained two-dimensional film given by the strain relation [26]:

$$\varepsilon_z = \varepsilon_{//} \frac{(-2\nu_{SiGe})}{(1-\nu_{SiGe})} \quad (1)$$

where $\varepsilon_z$ is the out-of-plane strain, $\varepsilon_{//}$ the biaxial in-plane strain and $\nu_{SiGe}$ is the Poisson ratio of the alloy deduced from Vegard's law. This result rules out the possibility of strain relaxation by defects and confirms the nominal strain value inside the layer. Similar measurements were performed in all unetched samples and also confirm the nominal strain state. Second, the same scans performed using the etched samples show no strain variation with respect to the unetched samples indicating the absence of etch induced relaxation by generation of defects. Kinematical simulations of the diffraction profiles were performed by a direct calculation of the scattering from a linear chain of atoms. The Cu Kα$_2$ line was incorporated by simultaneous simulation using a Kα$_2$ / Kα$_1$ intensity ratio of ½. The X-ray attenuation length at the (004) reflection for SiGe alloys measured here is of the order of 25 μm and was taken into account for the simulations. Given the relatively small thickness of the top alloyed film, no significant change of the diffraction profile after the etching is observed besides the integrated intensity reduction of the $Si_{0.40}Ge_{0.60}$ Bragg peak. For the coplanar geometry used, the integrated diffraction intensity for the thin strained film is proportional to the total volume of material. Since the diffracted peaks of interest sit in a non-linear background, it was necessary to qualitatively simulate the surrounding structures. The kinematical approach used was successfully employed to obtain a realistic background and to simulate the topmost strained layer intensity although it does not quantitatively hold for the substrate and buffer peaks [27]. Figures *2(c)-(d)* show a detailed view of the film peak measurements and simulation before and after etching. A fit without the top $Si_{0.40}Ge_{0.60}$ film performed to obtain the background is also shown as a dashed line. From the fits it is possible to deduce the etching rates for each sample with a statistical averaging over a ~ $1mm^2$ area. For sample A, the etching rate was found to be (4.0 ± 0.5) nm/min while sample B exhibits an etching rate of (4.9 ± 0.7) nm/min. Similar values are obtained by directly evaluating the area below the peaks after subtraction of the dashed line seen in Figure *2(c)-(d)*. The obtained etching rates are compatible within their error bars suggesting already that they are not significantly influenced by the strain state of the films. In order to corroborate this result, we perform systematic etching experiments on samples having the same composition but different strain states.

Prior to etching, we fabricate rectangular structures by photolithography followed by deposition of about 20 nm of Cr. A typical AFM scan of such a structure is shown in the inset of Figure *3(a)*. The as-defined mesas were then used as etch masks for the successive etching steps. Representative AFM linescans taken at the *same surface location* after various etching times are shown in Figure 3*(a)*. The etching rate is then simply deduced from the height of the profile and the etching time, taking into account the original Cr thickness. Figure 3*(b)* summarizes the etching rates for different samples with different strain states (compressive, tensile and relaxed) and Ge contents varying between 40 and 80%. It is obvious that the etching rates do not depend appreciably on the actual strain state in the films, i.e. for a given composition, they are compatible within 10-15 %. In contrast, the etching rates strongly depend on the Ge content $x$, increasing approximately exponentially with increasing $x$. Such a composition sensitivity was previously reported by Katsaros et al. [24] for relaxed SiGe films. The etching rates obtained from the X-ray diffraction measurements (Figure 2) over a much larger real space region are plotted as dashed lines in Figure 3*(b)* and corroborate those deduced from the AFM analyses. Finally, we perform a similar experiment by using a $H_2O_2$ solution to etch $Si_{0.20}Ge_{0.80}$ films with different strain states. The results are displayed in Figure 3*(c)*. Also in this case, the etching rate is not sensitive to the strain state in the films. This result, which was already expected from the disagreement between the calculated strain energy distribution and the experimentally observed etch profile in SiGe pyramids [11], is now clearly established.

For various applications, it might be important to know also whether the etching is isotropic or anisotropic. In order to answer this question, we processed circular ring structures on top of either Si(001) or $Si_{0.20}Ge_{0.80}$ relaxed buffers using photolithography and subsequent deposition of 100 nm Cr. Figure 4 shows SEM images of a typical ring structure defined on a Si(001) substrate by photolithography, Cr deposition and lift-off prior to (Figure 4*(a)*) and after wet chemical etching in a 2M KOH solution for two hours and subsequent removal of the underetched Cr layer in an ultrasonic bath (Figure 4*(b)*). The original ring structure evolves into a faceted octagon-like structure, because of the anisotropic etching rate of KOH [28,29]. The octagon sidewalls are tilted by about 50° with respect of the (001) surface, which is close to the {111} facet orientation (i.e. 54.7°). The latter facets are indeed known to be etched much slower than other crystal orientations [28,29]. The same ring structures were then defined on $Si_{0.20}Ge_{0.80}$ relaxed buffers. Figure 4*(c)* shows the ring structure after selective wet chemical etching in a $NH_4OH(1):H_2O_2$ (1) solution for 30 min. and subsequent removal of the underetched Cr layer. We can clearly see that the ring shape is preserved indicating that the etching is isotropic. A closer look to the ring sidewalls (not shown here) shows that no clear facets can be resolved in that case. Instead, the sidewalls appear rough, which may originate from the rough edges of the Cr rings. Finally, we apply the same procedure using a $H_2O_2$ solution. Figure 4*(d)* shows a typical ring structure after selective wet chemical etching in a $H_2O_2$ solution for 150 min. In this case, the underetched Cr layer was not removed. The etched regions are delimited by dashed lines, suggesting that the etching is also isotropic as for the $NH_4OH(1):H_2O_2$ (1) solution. Also in this case a closer look under the etched ring (not shown here) shows that no preferential facets develop during etching.

**4. Conclusion**

In conclusion, we have investigated the etching behavior of $NH_4OH(1):H_2O_2$ (1) and $H_2O_2$ solutions on both relaxed and biaxially strained $Si_{1-x}Ge_x$ films and we have observed that the etching rates are not affected by the actual strain state in the SiGe films.

We found that the etching rates of the $NH_4OH(1):H_2O_2(1)$ solution are primarily determined by the alloy composition. Moreover, the etching is isotropic for both etchants with no preferential facets developing during etching. Our results help to get a better understanding of SiGe wet chemical etching and pave the way towards the quantitative determination of three-dimensional compositional profiles of single SiGe/Si(001) islands.

**Acknowledgments**

The authors acknowledge Prof. Eric J. Mittemeijer and Dr. Udo Welzel (MPI-MF, Stuttgart) for the use of the X-ray rotating anode facility, Gerd Maier for technical support, J. Weber for assistance in the SEM characterization, G. Katsaros, M. Leite and G. Medeiros-Ribeiro for fruitful discussions and Prof. K. von Klitzing for his continuous support and interest. This work was supported by the BMBF (No. 03N8711) and the EU (No. 012150).

[27] For sample A, the buffer peak was obtained by simulating a 1 μm thick $Si_{0.20}Ge_{0.80}$ relaxed layer. For sample B, seven 1 μm thick relaxed layers with graded composition varying in steps of 5% Ge content were used for the simulation.

[28] Kendall D L 1975 *Appl. Phys. Lett.* **26** 195

[29] Seidel H, Csepregi L, Heuberger A and Baumgärtel H 1990 *J. Electrochem. Soc.* **137** 3612

**Figure captions:**

**Figure 1.** *(a)* Generic sample structure, *(b)* SIMS profile of sample A (tensile strained). The nominal layer structure is shown in the inset *(c)* SIMS profile of sample B (compressively strained). The nominal layer structure is shown in the inset (see text for details).

**Figure 2.** *(a)* X-ray longitudinal scan (θ-2θ) at the vicinity of the Si(004) reflection for sample A. Measurements performed at etched (open dots) and un-etched (solid dots) films are shifted on the logarithmic scale by multiplying the solid dots curve by a factor of 10. *(b)* Similar measurements performed with sample B. *(c)* and *(d)* show detailed zooms of the top strained layer diffraction peak. Kinematical simulations (discussed in the text) are represented by the solid (with top layer) and dashed (background without top layer) lines.

**Figure 3.** *(a)* AFM cross sectional profiles taken at the *same surface location* after Cr deposition prior to and after 40 min and 100 min etching in a $NH_4OH(1):H_2O_2(1)$ solution. The profiles are shifted vertically for clarity. A typical AFM scan of a rectangular mesa is shown in the inset. The dashed line defines the location of the cross sectional profiles. *(b)* Etching rate of a $NH_4OH(1):H_2O_2(1)$ solution versus etching time for different $Si_{1-x}Ge_x$ films with x = 0.4, 0.6 and 0.8 and different strain states: tensile strained (upward pointing triangle), compressive strained (downward pointing triangle) and relaxed (full circles). The dashed lines represent the upper and lower etching rates

deduced from the X-ray diffraction measurements shown in Figure 2*(c)* Etching rate of a $H_2O_2$ solution versus etching time for $Si_{0.20}Ge_{0.80}$ layers with different strain states: tensile strained (upward pointing triangle), compressive strained (downward pointing triangle) and relaxed (full circles). The detailed sample structure used for the etching experiments is shown in the inset.

**Figure 4.** Top view SEM image of a ring processed on a Si(001) surface prior to *(a)* and after wet chemical etching for two hours in a 2M KOH solution and subsequent removal of the underetched Cr layer *(b)*. The dashed lines indicate the position of the ring prior to etching. Top view SEM images of similar rings processed on a relaxed $Si_{0.2}Ge_{0.80}$ buffer after 30 min. wet chemical etching in a $NH_4OH(1):H_2O_2(1)$ solution and subsequent removal of the underetched Cr layer *(c)*, after 150 min. etching in a $H_2O_2$ solution *(d)*. The dashed lines delimit the edges of the etched regions.

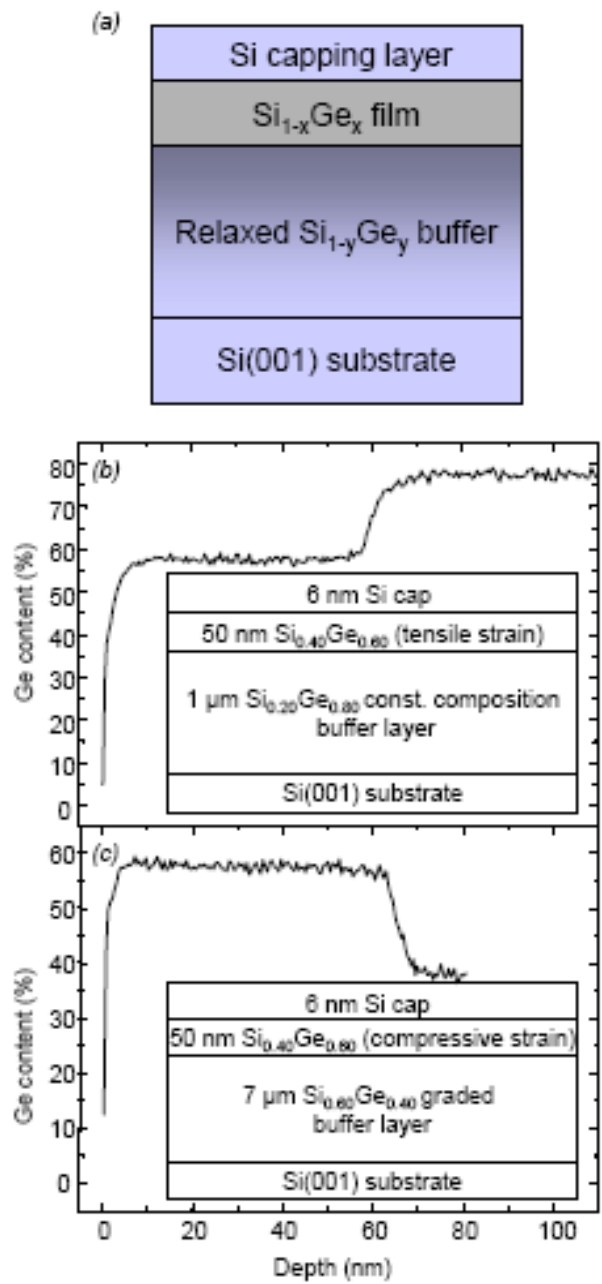

Fig. 1

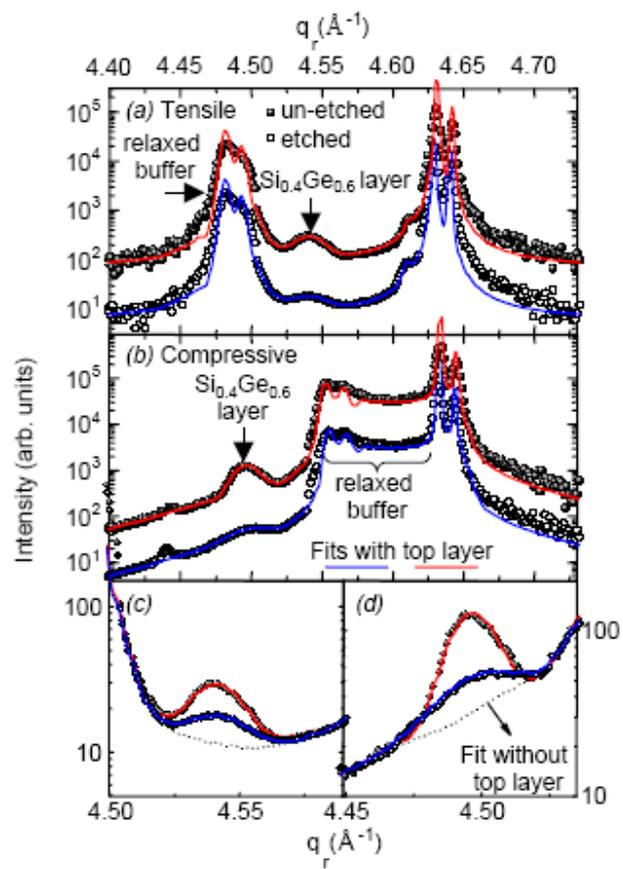

Fig. 2

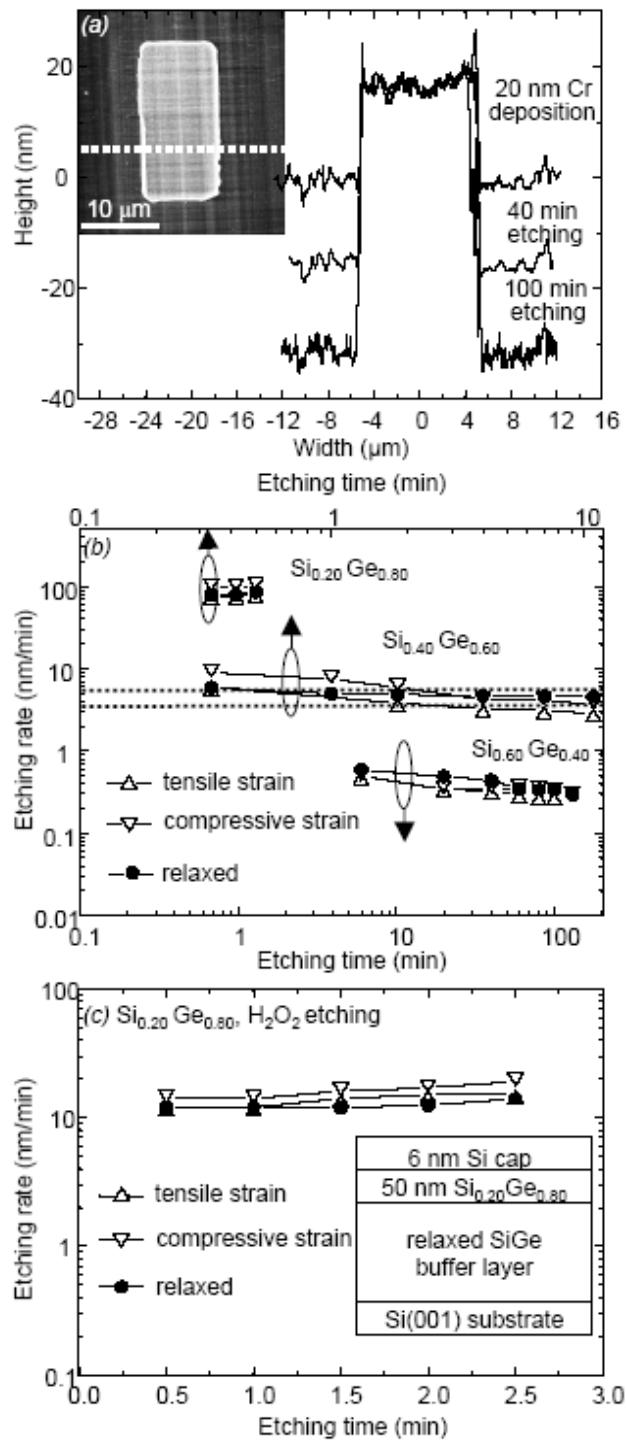

Fig. 3

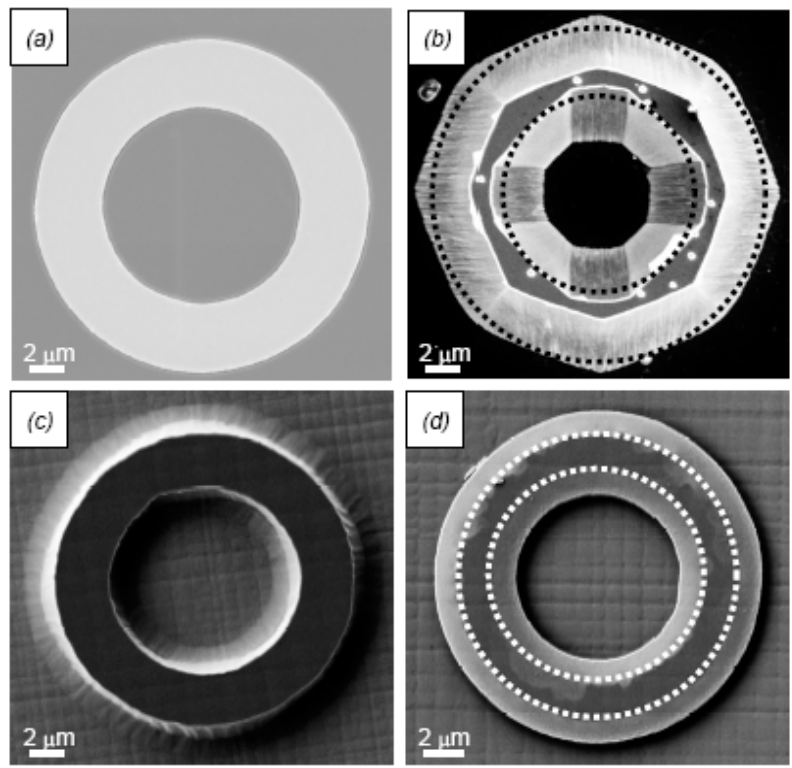

Fig. 4